# Spin-flop and antiferromagnetic phases of the ferromagnetic half-twist ladder compounds $Ba_3Cu_3In_4O_{12}$ and $Ba_3Cu_3Sc_4O_{12}$

M Kumar[1,2], S E Dutton[1,3], R J Cava[1] and Z G Soos[1]

[1]Department of Chemistry, Princeton University, Princeton NJ 08544, USA

[2] S. N. Bose National Centre for Basic Sciences, Block-JD, Sector-III, Salt Lake, Kolkata-700098

[3]Department of Physics, Cavendish Laboratory, University of Cambridge, JJ Thomson Avenue, Cambridge, CB3 0HE, UK

Email: soos@princeton.edu, manoranjan.kumar@bose.res.in

**Abstract**. The title compounds have dominant ferromagnetic (FM) exchange interactions within one-dimensional (1D) half-twist ladders of s = ½ $Cu^{2+}$ ions and antiferromagnetic (AFM) interactions between ladders, leading to ordered 3D phases at temperatures below 20K. Here we show that a microscopic 1D model of the paramagnetic (PM) phase combined with a phenomenological model based on sublattice magnetization describes the observed temperature and field dependent magnetism. The model identifies AFM, spin-flop (SF) and PM phases whose boundaries have sharp features in the experimental magnetization M(T,H) and specific heat $C_P(T,H)$. Exact diagonalization (ED) of the 1D model, possible for 24 spins due to special structural features of half-twist ladders, yields the magnetization and spin susceptibility of the PM phase. AFM interactions between ladders are included at the mean-field level using the field, $H_{AF}$, obtained from modelling the ordered phases. Isotropic exchange $J_1 = -135K$ and g-tensor g = 2.1 within ladders, plus exchange and anisotropy fields $H_{AF}$ and $H_A$, describe the ordered phases, and are almost quantitative for the PM phase.

## 1. Introduction

One-dimensional (1D) spin chains have been extensively studied, both experimentally and theoretically, and are good approximations for describing the magnetism of diverse inorganic and organic solids. Experimental studies have focused broadly on magnetic properties, instabilities, and 3D ordering transitions that eventually disrupt the 1D behaviour at low temperature [1,2]. Models of spin-1/2 chains are many-body quantum systems with interesting theoretical properties [3,4]. Both ferromagnetic (FM) and antiferromagnetic (AFM) exchange $J_1$ between nearest neighbours lead to frustration when there is AFM exchange, $J_2 = \alpha J_1 > 0$, between second neighbours. The spin-Peierls system [5] $CuGeO_3$, for example, displays spin-1/2 chains of $Cu^{2+}$ ions with $J_1 > 0$. Spin-1/2 chains of $Cu^{2+}$ with FM exchange $J_1 < 0$ have recently been identified in oxides [6–13] on either side of the quantum critical point [14], $\alpha_c = -1/4$, that is the exact boundary between a FM and a singlet ground state (GS) for isotropic exchange. Theory indicates the possibility of exotic ground states (GS) in FM/AFM chains with isotropic or anisotropic exchange [15–20]. Such chains include $Rb_2Cu_2Mo_3O_{12}$, $LiCuSbO_4$, $LiCuZrO_4$, $LiCuVO_4$ and $LiCu_2O_2$, and while they span a wide range of magnetic behaviour, they are characterized [13,15] by frustration $\alpha < \alpha_c$ and substantial $J_2$. By contrast, the compounds $Ba_3Cu_3In_4O_{12}$ and $Ba_3Cu_3Sc_4O_{12}$ discussed in this paper (and hereafter called In-334 and Sc-334) have small $\alpha \sim 0$ and dominant FM exchange within the chains [21].

The synthesis and structure of In-334 and Sc-334 have been reported recently [22]. Volkova et al. [23] and Koteswararao et al. [24] obtained the magnetization M(T,H) and specific heat $C_P(T,H)$ of In-334 and Sc-334, respectively, and Dutton et al. [21] studied both compounds. The magnetic lattice consists of half-twist ladders along the c axis instead of the more familiar $Cu^{2+}$ chains seen in other 1D oxides. Both compounds have FM exchange $J_1 < 0$ within the ladders and both undergo phase transitions [21,23,24] on cooling that clearly imply 3D interactions that are predominantly AFM. Experimental phase boundaries have been found, based on signatures in dM(T,H)/dH or $C_P(T,H)$ data. In zero field, the Néel transition to an AFM phase is at $T_N$ = 12.3K for In-334 and 15.2K for Sc-334. The saturation magnetic field at 2K is $\mu_0 H_{SAT}$ = 5.2T for In-334 and 8.0T for Sc-334. Microscopic modelling requires exchange constants between $Cu^{2+}$ spins in the same and adjacent half-twist ladders, as discussed by Volkova et al. [23] and calculated explicitly by Koteswararao et al. [24]. Neither group presented thermomagnetic fits, however. The nature of the ordered phases has not been identified, and microscopic modelling of ordered phases poses daunting challenges.

In this paper, we identify the ordered magnetic phases of In-334 or Sc-334 through the application of a phenomenological treatment [25] of sublattice magnetization used for crystals with spin-flop (SF) or AFM transitions. The method applies to crystals with FM subsystems that are not modelled at the microscopic level and describes the total M(T,H) by introducing two magnetic fields as parameters: an exchange field $H_{AF}$ that represents AFM interactions between subsystems and an anisotropy field $H_A$ that defines the subsystems' easy axis. We take the FM subsystems to be the half-twist ladders, solve a microscopic 1D model for the individual ladders and obtain the magnetic properties with mean-field corrections in $H_{AF}$.

Computational advances [3,4] have made it possible to model 1D systems in detail, especially those with spin-1/2. GS properties can be addressed by density matrix renormalization group (DMRG) methods [26], by Monte Carlo simulations [27] and by field theories [28] that can be checked against a limited number of exact results. Thermodynamic properties are still challenging, however, especially at the low temperatures that are most relevant for comparison with experiment. Bonner and Fisher [29] showed the power of exact diagonalization (ED) using N ~ 12 spins, about half the number that is accessible with current computers and methods [30]. Here we combine ED for a 1D model of half-twist ladders with mean field corrections, $H_{AF}$ and $H_A$, to account for 3D interchain interactions and the magnetic anisotropy respectively. Using this approach we extend the temperature range over which magnetic properties can be modelled and identify the phases in the experimentally observed M(T,H) phase diagram for In-334 and Sc-334. These complex magnetic phase diagrams provide a rare opportunity to check the consistency of parameters at different levels of theory for 1D magnetic systems.

The 1D magnetic lattice [21] of In-334 and Sc-334 is shown in figure 1a. The crystals are tetragonal with half-twist ladders running along the c-axis [21–24]. There are two distinct $Cu^{2+}$ sites: Cu1 forms square-planar $CuO_4$ complexes in the ab plane while Cu2 forms square-planar complexes that alternate in the ac and bc planes. The $CuO_4$ units in figure 1a are connected through corner-sharing oxygen. Each pair of Cu2 forms a rung normal to c, and successive rungs are twisted by $\pi/2$. The Cu-O bond lengths are essentially equal. The Cu1-O-Cu2 bond angles of 90.7° in In-334 and 87.8° in Sc-334 are consistent with (super) exchange $J_1 < 0$ between nearest neighbours. The half-twist ladder has equal $J_2 = \alpha J_1$ mediated by two O atoms between Cu2 in successive rungs. The spin Hamiltonian with isotropic exchange between first and second neighbours is

$$H(J_1,\alpha) = -|J_1|\sum_{p=1}(\vec{s}_{2pa}+\vec{s}_{2pb})\cdot\left[(\vec{s}_{2p-1}+\vec{s}_{2p+1})+\alpha(\vec{s}_{(2p+2)a}+\vec{s}_{(2p+2)b})\right] \qquad (1)$$

There is a single $J_1$ between Cu1 and Cu2 neighbours and $J_2$ is defined as exchange between successive Cu2 rungs as sketched in figure 1b. Since the spin $S_{2p} = s_{2pa} + s_{2pb}$ at each rung is conserved, $H(J_1,\alpha)$ reduces to sectors in which $S_{2p} = 1$ or 0.

The structure of half-twist ladders differs in several ways from $Cu^{2+}$ chains [5-13] with frustrated FM/AFM exchange and edge-sharing, coplanar $CuO_4$ complexes. The local xy-planes of $CuO_4$ complexes in figure 1a define three mutually orthogonal Cartesian frames. Since an unpaired electron in a Cu $d_{x2-y2}$ orbital typically has $g_\perp \sim 2.05$ in the plane and $g_\parallel \sim 2.20$ along z, the idealized half-twist structure has an isotropic average g-tensor, g ~ 2.1. Volkova et al. [23] report $g_\parallel = 2.15$ and $g_\perp = 2.08$ for In-334 based on electron spin resonance (ESR); the $CuO_4$ units are not quite planar and the crystal is tetragonal. The Zeeman interaction in an applied magnetic field is

$$H_Z = -g\mu_B \sum_{p=1} (\vec{s}_{2p-1} + \vec{S}_{2p}) \cdot \vec{H} \qquad (2)$$

where $\mu_B$ is the Bohr magneton. Half-twist ladders are modelled by $H(J_1,\alpha) + H_Z$ in the approximation of isotropic exchange and g-tensors. Additional exchange constants between spins in the same ladder can readily be included.

Finite size effects limit ED to 1D extended systems or to small 2D or 3D fragments. Adjacent ladders have nearby Cu1 and/or Cu2 spins with exchanges J' of unknown magnitude. We have previously shown [21] that neglecting J' is a good approximation for $T > 2T_N$. Here we treat the paramagnetic (PM) phase using an internal field $H_{AF}$ that stands for all J' between ladders. The effective field $H_e$ in Eq. 2 for the 1D model in Eq. 1 is then

$$H_e(T,H) = H - H_{AF} m(T,H_e)/\mu_B \qquad (3)$$

where $m(T,H_e)$ is the magnetization per spin. The relation $g\mu_B H_{AF} = zJ'$ between exchanges J' for spins $s_p$ and $s_{p'}$ in z neighbouring ladders follows from the mean-field (MF) approximation $s_{p'} \approx \langle s_{p'} \rangle = m(T,H_e)/g\mu_B$. We shall model the PM phase using Eqs. 1-3.

For ordered phases, we follow the method described by de Jongh and Miedema [25] for magnetic crystals with SF transitions. The sublattice magnetization is a phenomenological macroscopic description that holds for arbitrary spin. Some of the best realizations [25] of SF transitions are in [31] $MnCl_2(H_2O)_4$ with the s = 5/2 $Mn^{2+}$ ion and in [32] $GdAlO_3$ with the s = 7/2 $Gd^{3+}$ ion. Both have FM exchange in 2D layers whose direct modelling is very difficult and weak AFM exchange between layers. Willett, Gaura and Landee [33] review crystals with FM linear chains in terms of sublattice magnetization and discuss spin Hamiltonians in detail. AFM and SF phases are described by the magnetic fields $H_{AF}$ and $H_A$. Isotropic or almost isotropic g-tensors are required for SF transitions; large anisotropy $H_A$ leads instead to metamagnets [25,33].

It is well understood that models with isotropic exchange and g-tensors are a first approximation to the spin Hamiltonian. Since the magnitudes of additional terms tend to be poorly known, ED with multiple parameters becomes inefficient and, due to lower symmetry, more intensive computationally. It is particularly advantageous to conserve the total spin S, as done in Eqs. 1-3. On the other hand, ED provides direct comparisons with thermodynamic quantities and eigenstates for computing expectation values of additional terms of the spin Hamiltonian [33] such as anisotropic or antisymmetric exchange, g-tensor anisotropy, dipolar interactions between spins and hyperfine interactions.

The paper is organized as follows. In Section 2 we present the magnetic properties of half-twist ladders, discuss finite size effects at low temperature, and adapt SF relations for $H_{AF}$ and $H_A$ to polycrystalline samples. Section 3 starts with H-T phase diagrams of In-334 and Sc-334 based on magnetization and specific heat data [21]. The boundaries between the AFM, SF and PM phases are modelled using $H_{AF}$ and $H_A$. We find almost quantitative ED results in the PM phase for the spin susceptibility, magnetization and magnetic specific heat, as well as for recent ESR data [23]. The Discussion in Section 4 briefly addresses the choice of exchange constants and spin Hamiltonians for quasi-1D systems.

## 2. Magnetic properties of half-twist ladders
*2.1 Single ladder*

We set $|J_1| = 1$ in Eq. 1 and consider a half-twist ladder of $N = 3n$ spins with periodic boundary conditions in applied field $h = g\mu_B H$ with isotropic g in Eq. 2. Since the Zeeman and exchange Hamiltonians commute, the $h = 0$ energies are resolved into Zeeman levels $hS_z$ for $h > 0$. As noted above, the operators $S_{2p} = s_{2pa} + s_{2pb}$ for Cu2 spins in the same rung are constants of the motion in Eq. 1. We solve $H(J_1,\alpha)$ for all choices of $S_{2p} = 1$ or $0$ at each rung. Such partitioning reduces the dimensions from $2^n 2^{2n}$ states for 3n spins to $2^n 3^n$ in the largest sector with $S = 1$ at all rungs. ED becomes feasible up to $N = 24$ spins, whose largest sector is ten times smaller, $(3/4)^8 \sim 0.10$. The Hamiltonian for $N = 3n$ spins and $\alpha = 0$ is

$$H_N = \sum_{m=0}^{n} C(n,m) H_N(m)$$

$$H_N(m) = -\sum_{p=1}^{n} \vec{S}_{2p} \cdot \left(\vec{s}_{2p-1} + \vec{s}_{2p+1}\right)$$

(4)

Here C(n,m) is the binomial coefficient for having m out of n rungs with $S = 1$ and $H_N(m)$ has another sum for finite frustration $\alpha$. $H_N(n)$ is the ladder with $S = 1$ everywhere and contains the absolute GS with energy $E_0(n) = -n$ for 3n parallel spins, 4n exchanges $|J_1| = 1$ and $\alpha = 0$. $H_N(n-1)$ has one rung with $S = 0$, for example $S_{2n} = 0$, and GS energy $E_0 + 1$ as follows from deleting four $J_1$ at site 2n. Since any rung may be chosen to have $S = 0$, the $H_N(n-1)$ spectrum appears n times. Two rungs have S = 0 in $H_N(n-2)$ and they may be adjacent or not. The GS energy for $\alpha = 0$ is $E_0 + 3/2$ for adjacent rungs with S = 0 and $E_0 + 2$ when the rungs are not adjacent. It is straightforward to enumerate all unique distributions of S = 0 rungs and how many times each spectrum appears in the partition function. $H_N(0)$ with S = 0 everywhere is a Curie system of n spins Cu1.

We solve $H_N$, add $hS_z$ for h > 0, and construct the canonical partition function

$$Q_N(T,H) = \sum_{r=1}^{N/2}\sum_{S=0}\sum_{S_z=-S}^{S} \exp(-\beta(E_{rS} - hS_z)) \qquad (5)$$

Here $\beta = 1/k_B T$, $k_B$ is the Boltzmann constant, and $E_{rS}$ is the rth energy in sectors with total spin S. The magnetization per site is

$$m(T,H) = \frac{\partial \ln Q_N}{N\partial(\beta h)} = \frac{g\mu_B}{NQ_N}\sum_{r=1}^{N/2}\sum_{S=0}\sum_{S_z=-S}^{S} S_z \exp(-\beta(E_{rS} - hS_z)) \qquad (6)$$

The magnetic susceptibility is $\chi(T,H) = (\partial m/\partial h)_T$ and will be denoted in zero field as $\chi(T)$ per site and $\chi_M(T)$ per mole Cu. Other thermodynamic properties also follow directly from $Q_N$. The following results are for $\alpha = 0$ in Eq. 1. Calculations [21] with small frustration did not improve the fits significantly.

FM exchange leads to $\chi(T) > \chi_C = \beta g^2 \mu_B^2/4$, the Curie law for s = 1/2. The enhancement defines a correlation length $\xi(x)$ in terms of the reduced temperature $x = k_B T/|J_1|$,

$$\xi_N(x) = \chi_N(x)/\chi_C \qquad (7)$$

Finite N sets an upper bound on $\xi(x)$ since the GS has S = N/2. The T = 0 limit is $\xi_N(0) = (N+2)/3$ and $\xi(x)$ is constant until a state with smaller S is thermally populated. Figure 2 compares the size and temperature dependence of $\xi(x)$ of 334 ladders and (inset) Heisenberg FM chains. The dotted line is the exact solution of Yamada and Takahashi [34] as $x \rightarrow 0$. As expected on general grounds, $\xi_N(x)$ converges rapidly with N at high T where spin correlations are short ranged, while finite-size effects dominate at low T. ED up to N = 24 is quantitative for x > 0.2(T > $|J_1|/5k_B$) for ladders and fairly accurate to x = 0.1. ED for larger N is limited to low T with our computational resources because sectors with large S have smaller dimensions; a limited number of states can be found exactly in larger sectors. Dashed lines for N = 27 and 30 are to guide the eye. Finite-size effects in FM chains are less important at finite H because the field opens a gap.

*2.2 Mean-field treatment of the paramagnetic phase*

We approximate 3D exchange interactions by $H_{AF}$ in Eq. 3 and solve the 1D model $H_N$ with effective field $H_e$ in Eq. 2. The molar magnetization M(T,H) in the PM phase is modelled as

$$M(T,H) = N_A m(T, H_e) \tag{8}$$

with $H_{AF}$ taken from the ordered phases. The susceptibility $\chi_M(T,H)$ is a typical MF expression,

$$\chi_M(T,H) = \frac{\partial M(T,H)}{\partial H} = N_A \frac{\partial m}{\partial H_e}\frac{\partial H_e}{\partial H} = \frac{\chi_M(T,H_e)}{1+h_{AF}\chi(T,H_e)} \tag{9}$$

where $h_{AF} = g\mu_B H_{AF}$. The T dependence gives an estimate of the Néel temperature $T_N$,

$$\left(\frac{\partial m}{\partial T}\right)_H = \frac{\partial m}{\partial T} + \frac{\partial m}{\partial H_e}\frac{\partial H_e}{\partial T} = \frac{\partial m(T,H_e)}{\partial T}(1 - h_{AF}\chi(T,H_e)) \tag{10}$$

Since magnetization in constant field decreases with increasing temperature, the second expression must be negative and changes sign at $H_e = 0$, $T = T_N$. We obtain

$$1 = h_{AF}\chi(T_N) = \frac{g\mu_B H_{AF}\xi(x_N)}{4|J_1|x_N} \tag{11}$$

where $x_N = k_B T_N/|J_1|$. The MF approximation relates $H_{AF}$ in the AFM phase to $J_1$ in the PM phase; $\xi(x)/x$ is accurately given by ED with 30 spins for $x > 0.1$ and underestimated for $x < 0.1$.

The 3D contribution $h_{AF}\chi(T)$ also appears in Eq. 9 and becomes substantial on approaching $T_N$ from above. The GS spin degeneracy in zero field gives a residual entropy $S_{res} = k_B \ln N_A$ per mole. The degeneracy generates the $\xi(x)$ divergence [34] of the 1D Heisenberg chain in figure 2 and also a $T^{-1/2}$ divergence [35] in $C_P(T)/T$. The GS is not degenerate in a finite field, however, since it has $g\mu_B H$ lower energy than any other Zeeman level. The MF approximation underestimates 3D contributions near $T_N$. To see this, we performed ED on two parallel ladders with $J' = g\mu_B H_{AF}$ between sites 1 and 1', 2 and 2', ..., N and N'. The MF susceptibility $\chi_M(T)$ coincides with coupled chains at high T. Coupled chains have lower $\chi(T, J')$ at low T, a susceptibility maximum and $\chi(T, J') \to 0$ as $T \to 0$ because the GS is a singlet and there is an energy gap of order $J'$.

*2.3 Ordered phases*

We conclude this Section by summarizing and adapting expressions [25] for AFM and SF phase boundaries to polycrystalline samples. The anisotropy field $H_A$ specifies the easy axis along which sublattices in the AFM phase have opposite magnetization. An applied field H perpendicular to $H_A$ induces a net magnetization by canting the sublattices [25]

$$M_\perp(0,H) = \frac{H}{2H_{AF}+H_A}, \qquad H \leq H_\perp^{sat} \equiv 2H_{AF}+H_A \tag{12}$$

$M_\perp(0,H)$ increases linearly up to the SF/PM boundary where the saturation field orders the spins; $dM(0,H)/dH$ is constant up the $H_\perp^{sat}$ and vanishes at higher field. The 0K magnetization for H

parallel to $H_A$ is [25]

$$M_{\|}(0,H) = 0, \qquad H \leq H_{SF}$$
$$M_{\|}(0,H) = \frac{H}{2H_{AF} - H_A}, \qquad H_{SF} < H \leq H_{\|}^{sat} \equiv 2H_{AF} - H_A \qquad (13)$$

Since spin flips are activated processes for $H_{\|}$, there is no 0K magnetization up to the AFM/SF boundary where the magnetization flops perpendicular to $H_A$ in a first-order transition with discontinuous M and divergent dM/dH. The parallel saturation field is $2H_{AF} - H_A$, above which dM/dH = 0. The SF/AFM transition at 0K occurs at [25]

$$H_{SF} = \left(2H_{AF}H_A - H_A^2\right)^{1/2} \qquad (14)$$

These expressions have been successfully applied to single crystals with H parallel and perpendicular to the easy axis or the second easy axis in lower symmetry [25,33].

The simplest approximation for a polycrystalline sample is an isotropic distribution of easy axes at angle θ from the applied field H. We resolve H into a parallel and perpendicular component for each crystallite. Then $M_{\perp}(0,H)$ is along Hsinθ, normal to the easy axis. The component normal to the applied field cancels on summing over crystallites while the projection along the field is $Hsin^2θ$. We insert $Hsin^2θ$ into Eq. 12 and integrate over the distribution sinθdθ to obtain $2M_{\perp}(0,H)/3$. Crystallites with $θ < θ_{SF} = \cos^{-1}(H_{SF}/H)$ have flopped and hence contribute to $M_{\|}(0,H)$ as $Hcos^2θ$ in Eq.13. The integral over sinθdθ up to $θ_{SF}$ leads to

$$M_{\|}^{poly}(0,H) = \frac{1}{3}\frac{H\left(1-(H_{SF}/H)^3\right)}{(2H_{AF} - H_A)\left(1-(H_{SF}/H_{\|}^{sat})^3\right)}, \qquad H_{SF} \leq H \leq H_{\|}^{sat} \qquad (15)$$

The second factor in the denominator is inserted to give the correct saturation magnetization. As seen in figure 3, the magnetization of a polycrystalline sample is the sum of $M_{\|}(0,H)$ in Eq. 15 and $2M_{\perp}(0,H)/3$ in Eq.12. Discontinuous dM/dH at $H_{SF}$ and at the lower saturation field are due to $H_{\|}$ while finite dM/dH at H = 0 and discontinuous dM/dH at the higher saturation field are due to $H_{\perp}$. A polycrystalline sample combines features of parallel and perpendicular magnetization and has a range of SF transitions at $Hcosθ = H_{SF}$.

## 3. Modelling In-334 and Sc-334

*3.1 Phase diagram*

The phase boundaries of In-334 and Sc-334 shown in figure 4 were determined as follows [21]. Black squares are the saturation magnetic field $μ_0H_{SAT}$ down to 2K; red circles and light blue squares are two features in dM/dH. Blue squares and green triangles are respectively the field-dependent Néel temperature $T_N(H)$ obtained from the magnetic susceptibility $\chi_M(T,H)$ and specific heat $C_P(T,H)$. The close similarity of the two compounds is evident, and similar data have been reported for In-334 and Sc-334 in refs. [23] and [24], respectively.

To identify phases, we compare the magnetization at 2K in figure 5 with the phenomenological polycrystalline model at 0K given by Eqs. 15 and 12. The model has two parameters, $H_{AF}$ and $H_A$. The In-334 parameters were used in figure 3 to illustrate 0K features of $M(0,H)$ and $dM/dH$. The SF/AFM transition at $H_{SF} = 1.7T$ and saturation field $H_\perp^{sat} = 5.3T$ agree by construction, and $H_\parallel^{sat} = 3.9T$ is the dashed line in figure 4. The Sc-334 parameters in Table 1 generate similar $M(0,H)$ and $dM/dH$ curves that also support a spin-flop interpretation. We kept g = 2.1 in both fits even though the Sc-334 magnetic saturation is slightly less, g = 2.0, as seen in the left panel of figure 5; g = 2.1 is reported for Sc-334 in figure 3 of ref. [24]. It is instructive to view ordered phases as half-twist ladders in applied fields $H_{AF}$ that open gaps of $g\mu_B H_{AF} = 3.2K$ in In-334 and 4.9K in Sc-334. Spin flips are activated at low T while magnetization normal to $H_A$ is not. There is little change in $dM/dH$ vs. H curves at 5K or 8K, as seen in figure 4 of ref. [21].

The boundary of the PM phase in figure 4 is based on $\mu_0 H_{SAT}$ at low T and on the $C_P$ peak at $T_N(H)$ shown [21] in figure 6. The peak shifts to lower T with increasing H and the area under the peak decreases. Closely similar $C_P$ data with finer variations of H are shown for Sc-334 in figure 5b of ref. [24] and for In-334 in figure 3 of ref. [23], where $T_N = 12.7K$ is assigned not to the H = 0 peak at 12.3K but midway in the decrease above the peak. A second order transition marks the PM boundary with an expected [25] λ-anomaly at $T_N$. Well-resolved λ-anomalies with decreasing area are seen in $C_P(T_N(H))$ data on $MnCl_2(H_2O)_4$ single crystals [31]. We conclude that the phase boundaries of In-334 and Sc-334 are fully consistent with AFM, SF and PM phases.

The third model parameter in Table 1 is FM exchange $J_1$ that dominates the PM phase. We set $\alpha = 0$ in Eq. 1 after finding negligible improvement for small frustration. Before turning to the PM phase, we note that the MF approximation, Eq. 11, yields $T_N = 7.8$ and 11.4K for In-334 and Sc-334, respectively, reasonably close to the observed 12.3 and 15.2K.

Table 1. Magnetic parameters for In-334 and Sc-334

| Parameter | In-334 | Sc-334 |
|---|---|---|
| $J_1(K)$ | -135 | -135 |
| $H_{AF}(T)$ | 2.3 | 3.5 |
| $H_A(T)$ | 0.75 | 0.78 |

*3.2 Paramagnetic phase*

The PM phase permits quantitative modelling. Figure 7 shows the molar spin susceptibility [21] $\chi_M(T)$ of In-334 in a field of 0.1T. The $H_{AF} = 0$ and 2.3T curves are ED for N = 24 spins with $J_1 = -135$K, $\alpha = 0$ in Eq. 1 and g = 2.1 in Eq. 2. The MF approximation for AFM interactions between ladders extends the fit down to 20K (~ 0.15$|J_1|$) using $H_{AF}$ of the ordered phase. The Curie-Weiss law (inset) is satisfactory for T > 100K. We have neglected small temperature independent terms such as the diamagnetic susceptibility, $\chi_{dia} < 0$, and the Van-Vleck paramagnetism, $\chi_{VV} > 0$, of $Cu^{2+}$. ED with $H_{AF} = 0$ is satisfactory for T > 40K, where N = 18 calculations have converged according to figure 2. Finite-size effects come into play for T < 40K, with $\chi_M(T) > \chi_M(T, N = 24)$. As noted in Section 2, coupled ladders have $\chi(T, J') < \chi(T)$ with increasing $h_{AF}\chi(T)$. The dashed line in figure 7 has the same fractional correction as found using ED for coupled chains of 2N = 18. Better treatment of 3D interactions is needed close to $T_N$.

Larger $H_{AF} = 3.5$T for Sc-334 is consistent with higher $T_N = 15.2$K and lower $\chi_M(T)$ in figure 8 that is accurately fit down to 30K. The Curie-Weiss law in the inset holds for T > 100K. We have again neglected temperature independent contributions. Deviations due to $H_{AF}$ appear below 60K. As before, the dashed line is the minimal improvement for coupled chains with J' = $g\mu_B H_{AF}$.

Figure 9 compares the measured field dependence of the magnetization [21] in the PM phase of both materials with ED for N = 24 spins and parameters in Table 1. M(H,T) for In-334 is almost quantitative down to 20K. The 15K curve illustrates reduced finite-size effects in high field, where the fit improves. M(H,T) for Sc-334 is poorer but quantitative for T > 30K. The $\chi_M(T)$ curves in figures 7 and 8 ensure quantitative fits of the magnetization above 50K, where $M(T,H) = H\chi_M(T)$ holds for the M(T,H)/H data at 5 and 9T in figure 3 of ref. [21].

*3.3 Specific heat and electron spin resonance*

We return to $C_P(T,H)$ in the context of Heisenberg FM chains. $C_P(T,0)$ goes [35] as $T^{1/2}$ and has a broad maximum [28,36] around $k_BT \sim 0.3|J_1|$ that is of no concern here for $|J_1| > 100$K aside from noting the extended range of spin contributions. Since there is a finite energy gap for H > 0, $C_P(T,H)$ is exponentially small as T → 0. The field-induced gap leads [36] to $C_P(T,H)/T \sim a(H)$ at low T and H, and a(H) decreases with increasing field. In this regime, $C_P(T,H)$ has two maxima as a function of T that merge at large H as expected on general grounds.

The $C_P(T,H)/T$ peak in figure 6 fixes $T_N(H)$. Since $Q_N(T,H)$ in Eq. 5 does not refer to pressure or volume, we are computing only the T and H dependencies. Moreover, $C_P(T,H)$ for T

> $T_N(H)$ is dominated by phonons while $H > H_{SAT}$ is needed to extend the PM phase to $T < 10K$ where field-induced FM order may be a better description than AFM fluctuations. We suggested that $H_{AF}$ rationalizes the weak T dependence of $M(T,H)$ and $dM/dH$ for $T < T_N$. The same reasoning applies to the behaviour of $C_P(T,H)/T$ in figure 6, in particular the weak H dependence up to $H_{SAT}$ and the decrease for $H > H_{SAT}$. The 4T curve of In-334 and 6T curve of Sc-334 show $C_P(T,H)/T$ crossovers from the SF to the field-ordered PM phase, and $C_P(T,H)/T$ of In-334 decreases from 6 to 8T.

To extract the spin specific heat, the phonon contribution has to be estimated. The Debye expression $C_P/T = bT^2$ leads to the phonon lines in figure 6 based on data [21] to 30K with $b = 7.5 \times 10^{-4} J/K^4 molCu$ for In-334 and $4.8 \times 10^{-4} J/K^4 molCu$ for Sc-334. The In-334 value is close to the reported [23] $3b = 1.9 \times 10^{-3} J/K^4 mol$ based on $C_P(T,H)$ to $T = 300K$. The spin specific heat is obtained as usual from $\partial E(T,H)/\partial T$ at constant H using ED for 24 spins and $J_1 = -135K$. The spin plus phonon contributions in figure 6 are shown as dashed lines with $H = 6$ and 8T for In-334 and $H = 8T$ for Sc-334. The field-ordered PM phase is approximately fit without additional parameters.

The anisotropy field, $H_A = 0.75T$ for In-334 and $0.78T$ for Sc-334, has not appeared so far in our treatment of the PM phase. 350 GHz ESR of In-334 at 80K yields [23] a powder spectrum with $g_\parallel = 2.15$ and $g_\perp = 2.08$. The 0.5T difference in the resonance field the increases to 1.7T on cooling to 2K, as shown in figure 4 of ref. [23], with shifts of -0.8T for $g_\parallel$ and 0.4T for $g_\perp$. The g tensor is consistent with $g_\parallel$ along the half-twist ladder, the c axis in figure 1. Deviations from isotropic g are typically treated as perturbations.

We verified that the T dependence of the resonance field is proportional to $M(T,H)$ in a strong field of 10-12T that, as seen in figure 8, exceeds $H_{SAT}$ at 2K and completely aligns the spins. The magnetic field $B_c$ of parallel dipoles $g_\parallel \mu_B$ along c adds to the applied field, thereby reducing the resonance field $H_c$. The dipole field $B_{ab} = -B_c/2$ of spins polarized in the ab-plane opposes the applied field and increases the resonance field. In c-polarization, the dipolar field at Cu1 due to its four Cu2 neighbours is ~ 0.2T, while spins polarized in the ab plane lead to a field of ~ -0.1T at Cu1. Dipole-dipole interactions lead to sums that are conditionally convergent in 3D. Moreover, anisotropic exchange corrections [37] to $J_1$ also transform as dipoles. Dipole fields are traceless, however, and a resonance shift of -0.8T for H along c implies a resonance shift of 0.4T for H in the ab plane, as observed. Both $g_\parallel > g_\perp$ and dipoles point to c as the easy axis. The total difference at 2K between parallel and perpendicular resonance fields corresponds to the anisotropy field, $2H_A = 1.7T$, in excellent agreement with $H_A$ derived from the ordered phases.

## 4. Discussion

The magnetic structure of In-334 and Sc-334 (figure 1) contains corner-sharing $CuO_4$ complexes that form half-twist ladders along the crystallographic c axis. The half-twist geometry results in a different magnetic exchange pathway than in other 1D $Cu^{2+}$ chains with FM $J_1$. [6–13] Unlike the edge sharing Cu polyhedral chains with significant $J_2$ interactions and finite frustration α, In-334 and Sc-334 have dominant FM interactions modelled by $J_1 = -135K$ and α = 0 at high T. [21] Extending this result, we have included weak AFM exchange between the 1D ladders at the mean-field level through an internal field $H_{AF}$ in Eq. 3, thereby retaining the computational advantages of a 1D model. We have identified the magnetic phases in the experimentally determined M(H,T) phase diagrams in figure 4 as AFM and SF phases by adapting to polycrystalline samples a phenomenological treatment of sublattice magnetization in crystals. The anisotropy field $H_A$ and exchange field $H_{AF}$ between spins in different half-twist ladders successfully describes the thermomagnetic signatures observed in the ordered phases.

$H_{AF}$ and $H_A$ are phenomenological parameters [25] for crystals with FM subsystems, here taken to be the half-twist spin ladders. We comment below on the implications of combining phenomenological and microscopic modelling of In-334 and Sc-334. The parameters $J_1$, $H_{AF}$ and $H_A$ in Table 1 describe both ordered phases and, in some detail, the absolute values of the magnetization and spin susceptibility of the PM phase. Three parameters account for the basic temperature and field dependent magnetic properties of the 334 compounds and provide a baseline for future studies. More detailed experimental magnetic characterization of In-334 or Sc-334 would require the use of single crystals. FM exchange within the half-twist chains, $J_1$, is clearly central to the high temperature PM phase, where $H_{AF}$ and $H_A$ are secondary. Only $H_{AF}$ and $H_A$ appear explicitly in the description of the low temperature ordered phases, although $J_1$ enters in the estimate of $T_N$. We have not modelled the Néel transition, magnetism below 2K, or properties related to other terms of spin Hamiltonians.

More complex spin Hamiltonians can readily be proposed for In-334 or Sc-334; they illustrate the trade-offs between accuracy, number of parameters, and computational considerations. For example, reasonable $\chi_M(T)$ can be obtained [21] without $H_{AF}$ for various combinations of $J_1$ and small frustration, $0 > α > -0.06$, but finite $H_{AF}$ with α = 0 and $J_1 = -135K$ clearly improves agreement with experiment, as seen in Figs. 7-9. Modelling the ordered phases specifies the value of $H_{AF}$ in the PM phase, and of $H_A$ for ESR shifts. The MF estimate of $T_N$ in Eq. 11 depends on α, but from our perspective it is premature to invoke α as a free parameter in this context. For the same reason, $\chi_M(T)$ in figures 7 and 8 is due entirely to spins modelled by Eqs. 1-3. Fits based on $\chi_M(T) + \chi_0$ can only improve matters if the temperature independent term,

$\chi_0 = \chi_{dia} + \chi_{VV}$, is adjustable to some extent; the Van Vleck contribution is not known accurately, while $\chi_{dia}$ can be estimated using Pascal's constants. The estimated $\chi_0$ are small, -7 x $10^{-5}$cm$^3$/molCu for [24] Sc-334 and -9 x $10^{-5}$cm$^3$/molCu for [23] In-334. A combined fit would change $J_1$ slightly.

Exchange constants J are difficult to assign or compute because they are small on the scale of electronic energies and depend on both direct overlap between magnetic centres and through-bond pathways. We considered $J_{22} > 0$ in figure 1b between Cu2 in the same rung but did not find improved fits. Volkova et al. [23] and Kotewararao et al. [24] consider other exchanges such as (in the notation of figure 1b) $J_{22}$ and $J_{11}$ between adjacent Cu1-Cu1 that could be included in ED. They also suggest small exchanges between Cu in adjacent half-twist ladders. Table 2 of ref. [24] lists three sets of calculated exchanges for different on-site repulsion U, including -$J_1$ = 12.4meV = 144K, which is close to our fitted value. However, ED with this $J_1$ and other calculated exchanges within a ladder does not reproduce $\chi_M(T)$. Since finite-size effects are seen from figure 2 to be small for $k_BT/|J_1| > 0.3$, or T > 40K, proposed exchange constants within ladders can be tested directly. Volkova et al. [23] have suggested an exotic GS for In-334 at low T based on a 3D Shastry-Sutherland model [38] with three orthogonal dimers, the twisted Cu2Cu2 rungs of the ladder and Cu1Cu1 neighbours. Further experimental data below 2K will be needed to characterize the GS fully. Two points are worth noting: first, ED can be performed for postulated exchanges within a ladder and second, fits of magnetic data in the PM phase set fairly stringent limits on additional exchanges within or between ladders.

In summary, we have identified the antiferromagnetic and flip-flop phases in the M(H,T) phase diagram of Ba$_3$Cu$_3$In$_4$O$_{12}$ and Ba$_3$Cu$_3$Sc$_4$O$_{12}$ at low temperature. Transitions at 1.7 and 2.2T, respectively, lead to the magnitudes of the exchange and anisotropy fields H$_{AF}$ and H$_A$ in Table 1. The transition at T$_N$(H) is also consistent with the model. The paramagnetic phase is modelled as a half-twist ladder with isotropic exchange $J_1$ between Cu1 and Cu2 neighbours. Exact diagonalization accounts quantitatively for the magnetic susceptibility and magnetization at high temperature. A mean-field treatment of exchange interactions between ladders extends the fits almost to T$_N$ using H$_{AF}$ from the ordered phase. The three parameters $J_1$, H$_{AF}$ and H$_A$ provide a consistent picture and a starting point for more detailed studies of the magnetic interactions in these compounds.

**5. Acknowledgments**: This work was largely performed at the TIGRESS high performance computer centre at Princeton University, which is jointly supported by the Princeton Institute for Computational Science and Engineering and the Princeton University Office of Information

Technology. We thank the National Science Foundation for partial support of this work through the Princeton MRSEC (DMR-0819860), and DOE grant DEFG02-08ER46544 for support of S.E.D. and R.J.C.

**Figures**

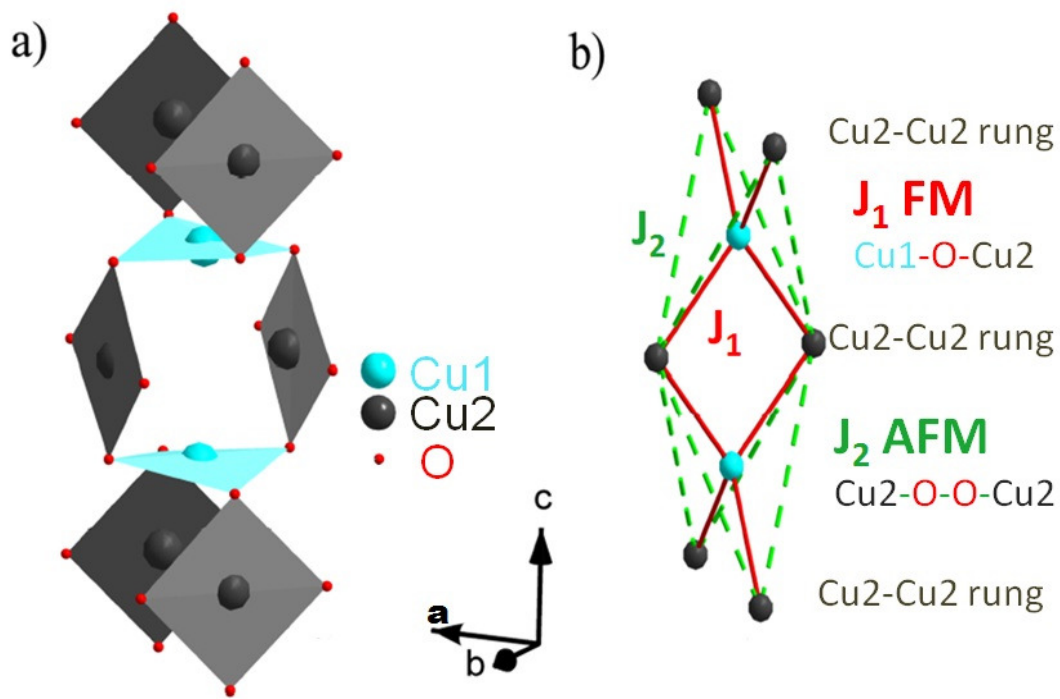

Figure 1: Magnetic lattice of half-twist ladders: (a) Corner sharing square-planar $CuO_4$ complexes with Cu1 along the ladder, the crystallographic c axis, and Cu2 rungs in the ab plane, successively twisted by $\pi/2$. (b) Nearest-neighbour FM exchange $J_1$ between Cu1 and Cu2 with Cu-O-Cu bond angle close to $\pi/2$ and AFM exchange $J_2$ mediated by two O atoms between Cu2 in successive rungs.

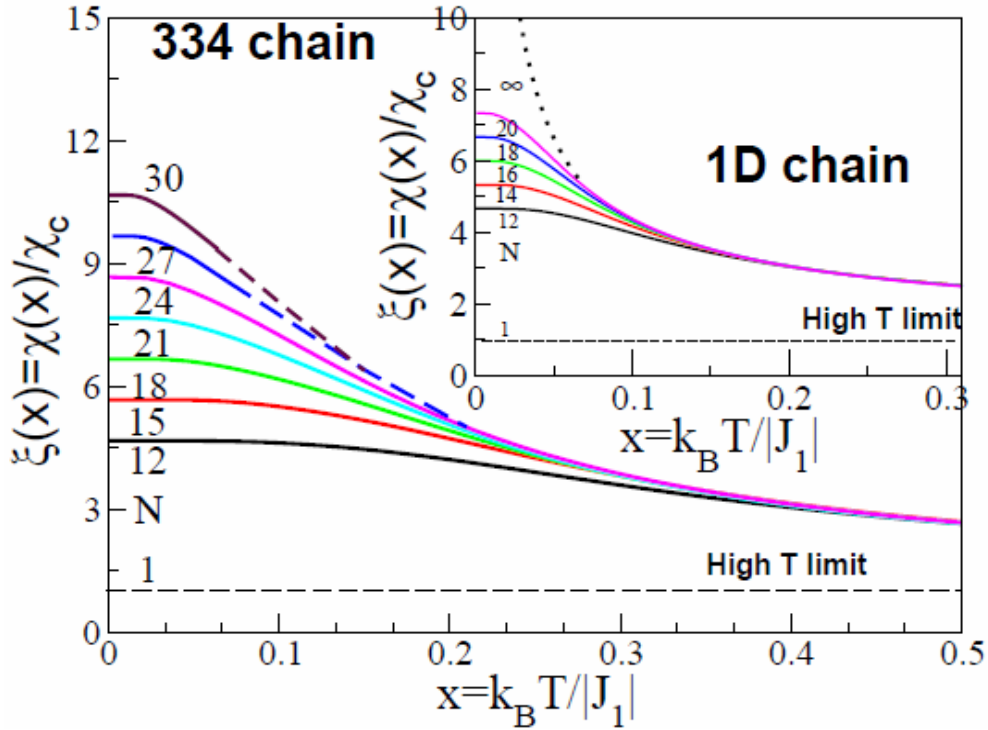

Figure 2. Correlation length $\xi_N(x)$, Eq. 7, of a half-twist ladder with $\alpha = 0$ in Eq. 1 and (inset) a 1D Heisenberg FM chain as a function of the reduced temperature $x = k_BT/|J_1|$. The infinite chain is from ref. [34]. Dashed lines are to guide the eye.

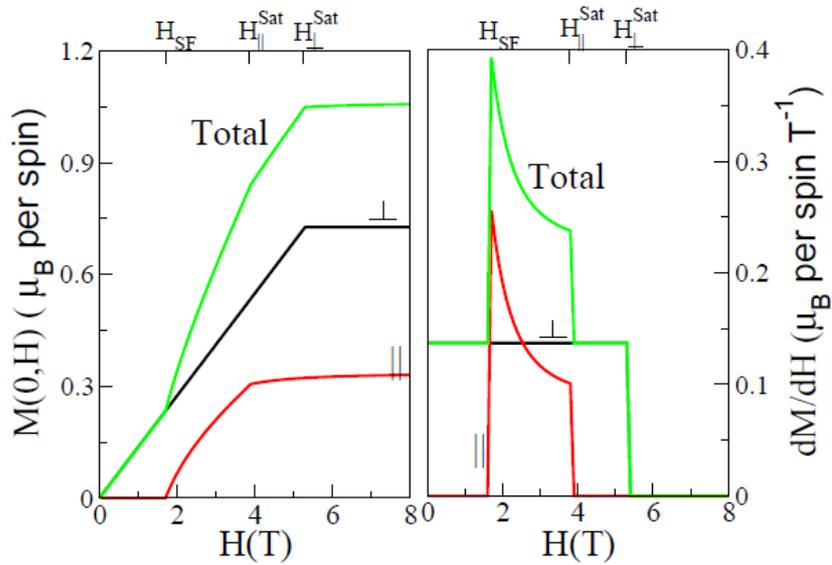

Figure 3. 0K magnetization $M(0,H)$ and $dM/dH$ vs. applied field H of a polycrystalline sample with spin-flop transition at $H_{SF} = 1.7T$ and saturation fields $H_\parallel^{sat} = 3.9T$ and $H_\perp^{sat} = 5.3T$ for H parallel and perpendicular to the easy axis. $M(0,H)/\mu_B$ saturates at $g/2$, $g = 2.1$; $dM/dH$ is discontinuous at $H_{SF}$ and at the saturation fields.

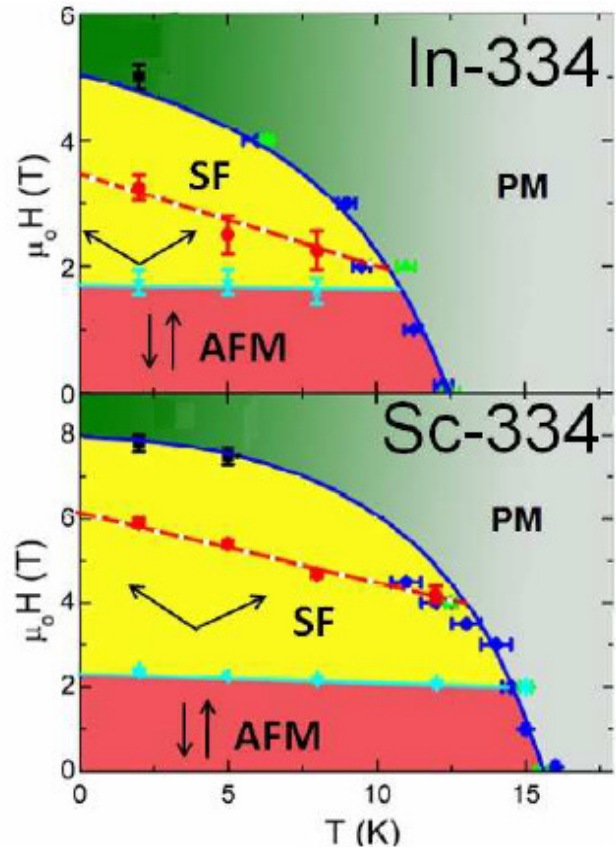

Figure 4. Phase diagram of (a) In-334 and (b) Sc-334 as a function of magnetic field, $\mu_0 H$, and temperature, T. Experimental boundaries are from ref. [21]; symbols refer to measurements discussed in the text.

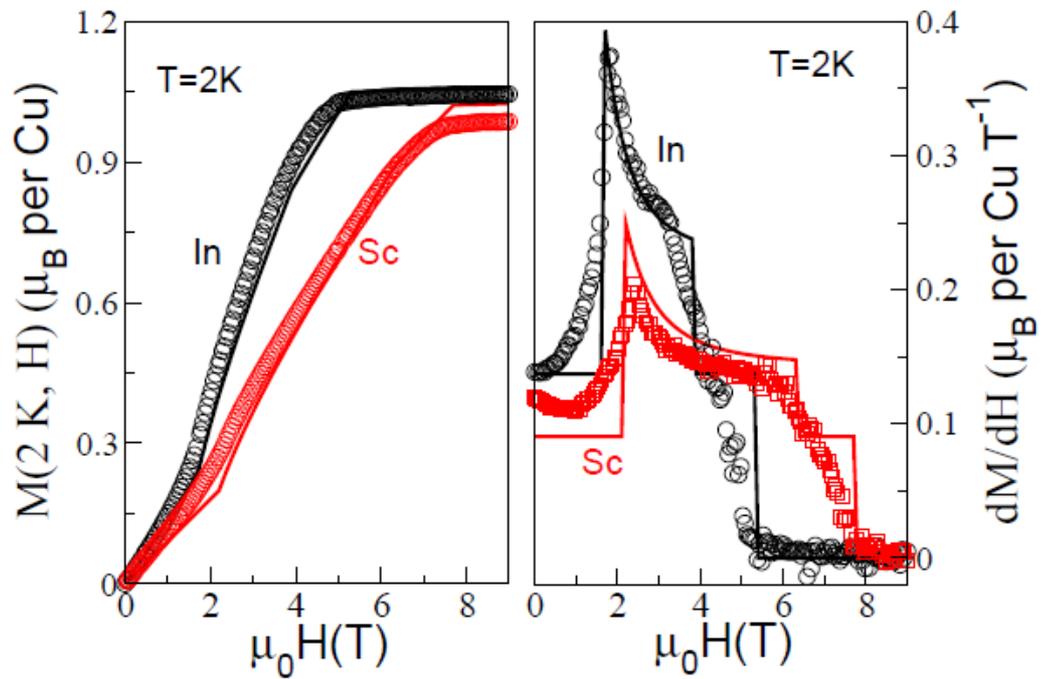

Figure 5. Lines are M(0,H) and dM/dH vs. applied field H based on Table 1 and Eqs. 15 and 12. Points are 2K data on polycrystalline In-334 and Sc-334 from ref. [21].

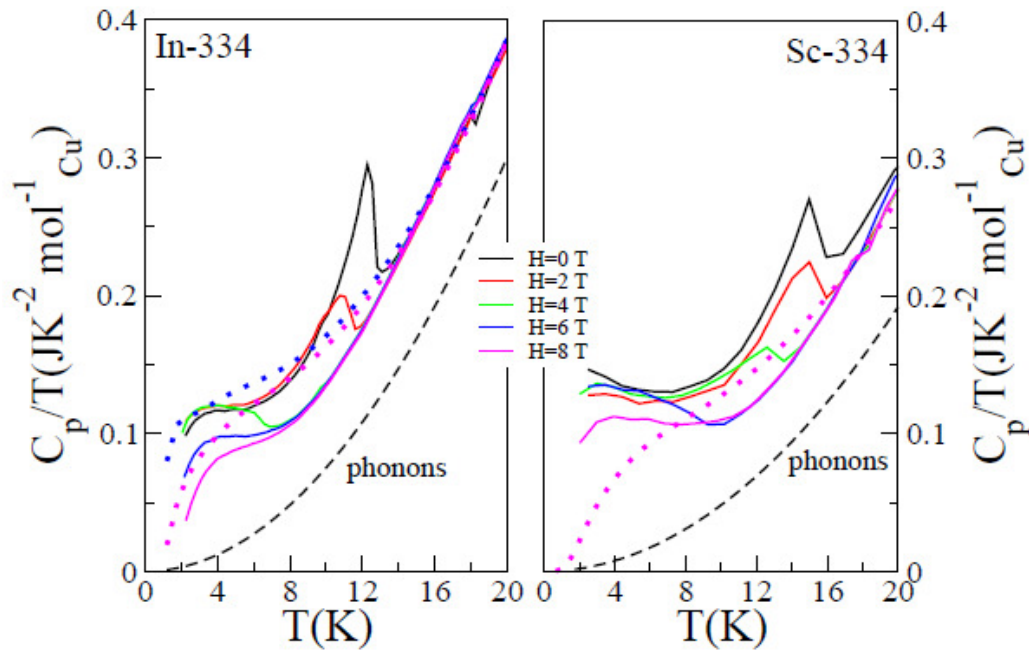

Figure 6. Lines show the temperature dependence of $C_P(T,H)/T$ at the indicated fields for In-334 (left panel) and Sc-334 (right panel), data from ref. [21]. The peak at the Néel temperature $T_N(H)$ is the boundary of the PM phase. See text for the phonon contribution. Dotted lines are calculated for 24 spins in applied fields of 6T and 8T for In-334 and 8T for Sc-334.

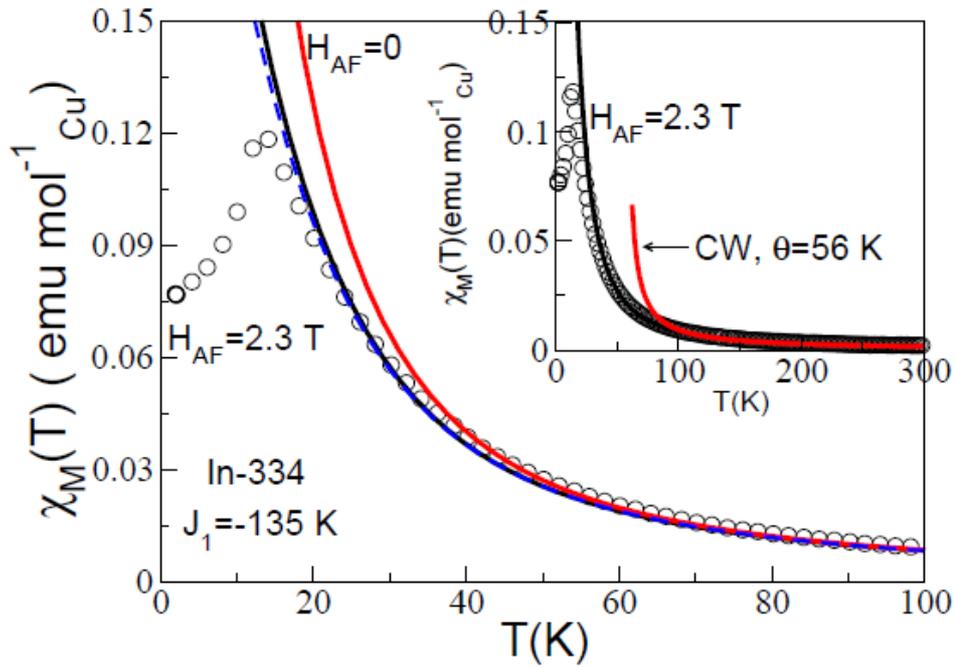

Figure 7. Temperature dependence of $\chi_M(T)$, the molar spin susceptibility of In-334 in the PM phase, data from ref. [21]. Lines are ED results for N = 24 spins with g = 2.1 and parameters in Table 1. $H_{AF}$ is AFM exchange between half-twist ladders. The Curie-Weiss fit in the inset has FM Weiss constant of 56K. Dashed lines are corrections for coupled chains discussed in the text.

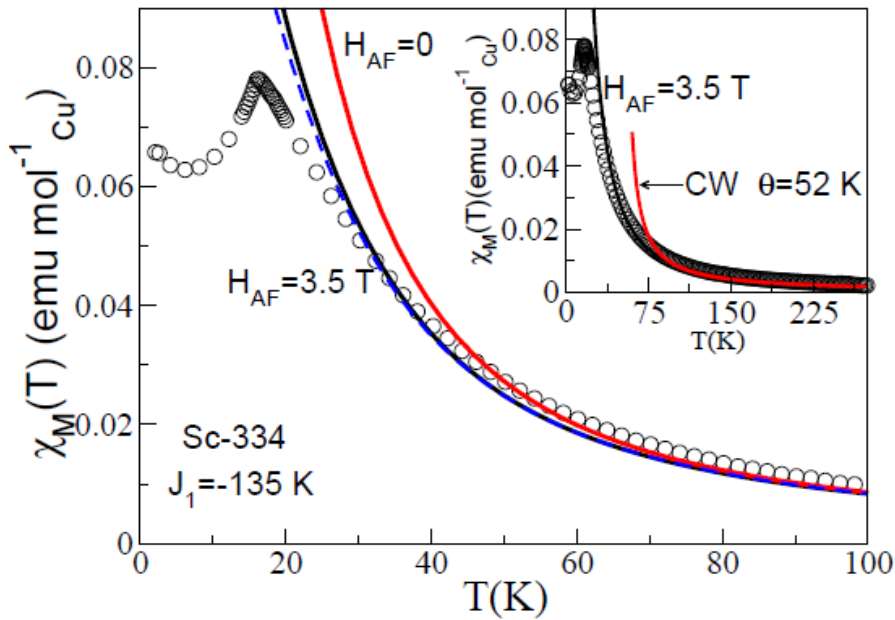

Figure 8. Same as Fig. 7 for Sc-334 with $H_{AF}$ = 3.5T and FM Weiss constant of 52K.

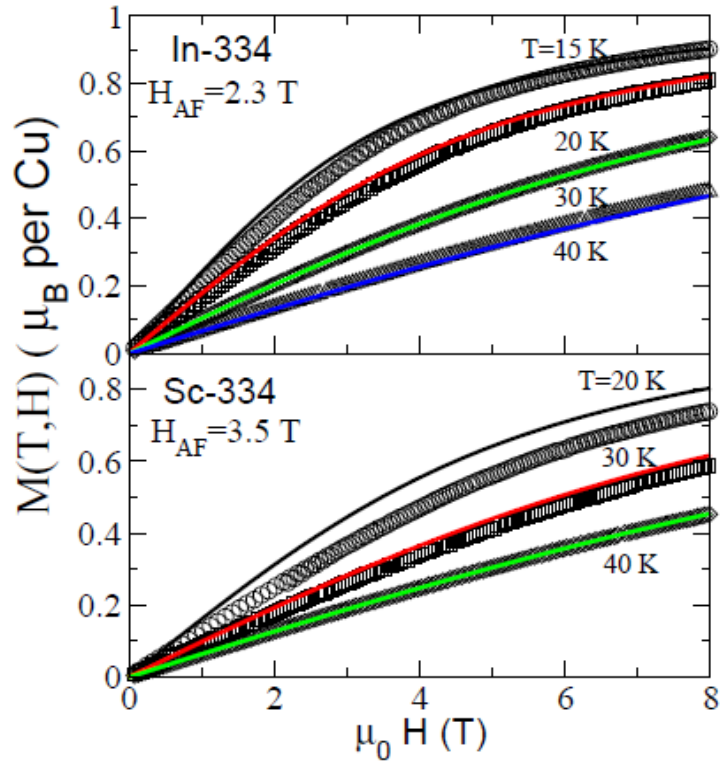

Figure 9. Field dependence of the magnetization M(T,H) of In-334 (upper panel) and Sc-334 (lower panel) with data from ref [21]. Lines are ED for 24 spins with g = 2.1 and parameters in Table 1.